# Epitaxial Antiperovskite/Perovskite Heterostructures for Materials Design

Camilo X. Quintela,[1] Kyung Song,[2] Ding-Fu Shao,[3] Lin Xie,[4] Tianxiang Nan,[1] Tula R. Paudel,[3] Neil Campbell,[5] Xiaoqing Pan,[7] Thomas Tybell[8], Mark S. Rzchowski,[5] Evgeny Y. Tsymbal,[3] Si-Young Choi,[6*] Chang-Beom Eom[1*]

**Engineered heterostructures formed by complex oxide materials are a rich source of emergent phenomena and technological applications. In the quest for new functionality, a vastly unexplored avenue is interfacing oxide perovskites with materials having dissimilar crystallochemical properties. Here, we propose a unique class of heterointerfaces based on nitride antiperovskite and oxide perovskite materials as a new direction for materials design. We demonstrate fabrication of atomically sharp interfaces between nitride antiperovskite $Mn_3GaN$ and oxide perovskites $(La_{0.3}Sr_{0.7})(Al_{0.65}Ta_{0.35})O_3$ (LSAT) and $SrTiO_3$. Using atomic resolution imaging / spectroscopic techniques and first-principle calculations, we determine the atomic-scale structure, composition, and bonding at the interface. The epitaxial antiperovskite / perovskite heterointerface is mediated by a coherent interfacial monolayer that interpolates between the two anti-structures. We anticipate our results to be a major step for the development of functional antiperovskite/perovskite heterostructures, combining their unique characteristics such as topological properties for ultra low power applications.**

[1] Department of Materials Science and Engineering, University of Wisconsin-Madison, Madison, Wisconsin 53706, USA. [2] Department of Materials Modeling and Characterization, KIMS, Changwon 51508, South Korea. [3] Department of Physics and Astronomy & Nebraska Center for Materials and Nanoscience, University of Nebraska, Lincoln, Nebraska 68588, USA. [4] National Laboratory of Solid State Microstructures and College of Engineering and Applied Sciences, Nanjing University, Nanjing, Jiangsu 210093, People's Republic of China. [5] Department of Physics, University of Wisconsin-Madison, Madison, Wisconsin 53706, USA. [6] Department of Materials Science and Engineering, POSTECH, Pohang 37673, South Korea. [7] Department of Chemical Engineering and Materials Science, University of California-Irvine, Irvine, CA 92697, USA. [8] Department of Electronic Systems, Norwegian University of Science and Technology, Trondheim 7491, Norway;

*Corresponding author. Email: eom@engr.wisc.edu, youngchoi@postech.ac.kr

## Introduction

Complex oxide materials, and in particular heterostructures formed from them, are a rich source of emergent phenomena and technological applications (1-3). Interfacing oxide perovskites with materials having dissimilar crystallochemical properties and functionalities (4), are likely to vastly expand the range of interfacial phenomena and applications. However, stabilizing such heterostructures with the chemical and structural quality required to promote the desired functionality is challenging when the constituting materials are non-isostructural having large geometrical and chemical strains (5-6). We propose a unique class of heterointerfaces based on nitride antiperovskite and oxide perovskite materials as a new direction for materials design.

Antiperovskite materials are intermetallic compounds with perovskite crystal structure (space group $Pm\overline{3}m$, No. 221) but with anion and cation positions interchanged in the unit cell (7). Similar to their oxide-perovskite counterparts, antiperovskite materials possess a variety of tunable physical properties including superconductivity, itinerant antiferromagnetism, giant magnetoresistance, large magnetovolume effects, and topological electronic behavior (8-15). Among antiperovskite materials, transition metal (TM)-based nitride compounds ($M_3X$N; $M$: TM, $X$: metallic or semiconducting element) are particularly interesting as their physical behaviors are remarkably sensitive to external perturbations such as magnetic fields, temperature, or pressure (14-19). This is mainly due to the strong spin-lattice coupling characteristic of $M_3X$N compounds. With such a correlated physical background, the development of epitaxial $M_3X$N heterostructures provides an ideal platform for tuning the properties of $M_3X$N with the proper choice of materials and design. In this context, $AB$O$_3$ oxide perovskites are unrivaled material systems to interface with $M_3X$N nitride antiperovskites as both compounds have analogous perovskite-type crystal structure with comparable lattice constants, affording good epitaxial match along any common crystallographic direction and should thus promote epitaxial growth. This enables the use of strain engineering to tune the behavior of $M_3X$N materials. Additionally, the wide variety of physical properties of $AB$O$_3$ compounds can be used as external triggers to tune the functionality of antiperovskite materials, allowing the development of multifunctional artificial materials and devices, such as recently proposed for heterostructures between $Mn_3GaN$ and oxide ferroelectric and piezoelectric perovskites (16-18).

To exploit this potential, it is first necessary to understand at the atomic level the interfacial structure and chemistry between nitride antiperovskite and oxide perovskite materials in order to promote a bridging structure allowing for epitaxy. From the crystallographic perspective, the atomic configuration at the interface between these two anti-structures is not obvious. As illustrated in Fig. 1A, $M_3X$N antiperovskite and $AB$O$_3$ perovskite compounds show reversed anion and cation positions in the unit cell. This distinctive difference leads to different considerations for interfaces between $M_3X$N and $AB$O$_3$ materials than that between two perovskite or two antiperovskite materials (Fig. 1B). The $AB$O$_3$ perovskite structure can be described as alternating mixed cation-anion $A$O and $B$O$_2$ layers along the $[001]_{\text{perovskite}}$ ($[001]_{\text{P}}$) direction of the unit cell, and only two trivial interfacial configurations are physically stable: A'O/BO$_2$ and B'O$_2$/AO (Fig. 1C) for interfaces formed between two different $AB$O$_3$ and $A'B'$O$_3$ compounds. Using the same analogy, nitride $M_3X$N antiperovskites can be viewed as a stacking of alternating $MX$ and $M_2$N layers along the $[001]_P$ direction. As illustrated in Fig. 1D, the number of hypothetical simplest possible interfacial configurations between $M_3X$N antiperovskite and $AB$O$_3$ perovskite materials doubles to four depending on the termination of the $AB$O$_3$ perovskite: $M_2$N/$B$O$_2$, $MX$/$B$O$_2$, $M_2$N/$A$O and $MX$/AO. However, a subsequent consideration is the chemical bonding at the interface between nitride antiperovskite and oxide perovskite materials to promote epitaxial growth. Contrary to oxide perovskites, which have predominantly ionic bonding, nitride antiperovskites generally show metallic/covalent chemical bonding. In this context, developing a strategy to properly interface nitride antiperovskites with oxide perovskite materials can facilitate the emergence of interfacial hybridization interactions and hence new interfacial properties and functionalities not achievable in more conventional oxide/oxide interfaces, opening a new path in the search for emergent behavior linked to interfacial phenomena (4).

Sparked by the quest for fundamental understanding of the nitride-antiperovskite/oxide-perovskite interface, we establish high-quality epitaxial $Mn_3GaN$ films on (001)-oriented LSAT and $SrTiO_3$ single-crystal substrates as paradigms of $M_3X$N/$AB$O$_3$ interfaces. Using a combination atomic resolution STEM, electron energy loss spectroscopy (EELS), and energy-dispersive x-ray spectroscopy (EDS) techniques, and density functional theory study, we unveiled the mysterious interfacial structure of $Mn_3GaN$/LSAT and $Mn_3GaN$/$SrTiO_3$ on an atomic scale. We investigated both the stability and the mechanism of nucleation of the observed interface using first principles calculations. For simplicity, the manuscript focus on the $Mn_3GaN$/LSAT interface, and additional information, including experimental data regarding the $Mn_3GaN$/$SrTiO_3$ interface and materials and methods, is presented in the Supplementary Information.

**Results**

Fig. 2 summarizes the X-ray diffraction (XRD) structural characterization for a 60 nm thick $Mn_3GaN$ film grown on a (001) LSAT substrate. The epitaxial growth and single-phase structure of the films was monitored using in-situ reflection high-energy electron diffraction (RHEED) and confirmed through symmetric θ-2θ XRD measurements by the observation of only the (00*l*) reflections (Fig. 2A). In Fig. 2B, a representative θ-2θ XRD scan taken around the (002) LSAT substrate peak is depicted. The presence of Kiessig fringes surrounding the $Mn_3GaN$ (002) reflection indicates high crystalline quality of the film and a pristine interface, and corroborate the narrow 0.035° full-width at half-maximum (FWHM) value of the rocking curve for $Mn_3GaN$ (002) (Fig. 2C). Decreasing film thickness results in an improvement of the crystallinity, reaching films with FWHM values as low as 0.023°. An in-plane cube-on-cube epitaxial relationship between $Mn_3GaN$ and substrate was confirmed by off-axis azimuthal ϕ-scan around the (022) reflection (Fig. 2D). From x-ray reciprocal space mapping (RSM) measurements centered in the asymmetrical (-113) LSAT peak (Fig. 2E), the out-of-plane ($a_\perp$) and in-plane ($a_\parallel$) lattice constants were determined at $a_\perp = 3.90 \pm 0.01$ Å and $a_\parallel = 3.92 \pm 0.01$ Å, close to the bulk lattice constant of $a = 3.898$ Å (20).

To investigate the structure and chemical composition of the $Mn_3GaN$/LSAT interface a combination of atomic-resolution STEM, EELS, and EDS techniques were employed. For additional analyses, including data for the $Mn_3GaN/SrTiO_3$ interface, please see the Supplementary Information. In Fig. 3A, an atomic-resolution high angle annular dark field (HAADF) STEM image taken along the [100] zone axis of LSAT is depicted. The image displays an atomically sharp interface and further corroborates the high crystalline quality of the films. In Fig. 3B, a magnified HAADF-STEM image close to the epitaxial $Mn_3GaN$/LSAT interface, is shown, overlaid with the cation positions as determined by this study. The atomic resolution EDS and EELS analyses (Fig. S1 and Fig. S2) and HAADF-STEM intensity profiles (Fig. 3C) demonstrate that the LSAT substrate termination is $(Al_{0.65}Ta_{0.35})O_2$ ($BO_2$ termination) and thus implies that the $Mn_3GaN$ termination at the interface is expected to be compatible with $Mn_2N$. However, the first $Mn_3GaN$ interfacial monolayer (labeled as layer 1 in Fig. 3C) exhibits a pattern of alternating bright and dark spots, indicative of compositional and/or structural reconfigurations at the interface.

Albeit the above data shows that it is possible to epitaxially interface an antiperosvkite with a perovskite, to detail the bridging structure thorough atomic-resolution EDS and EELS experiments were performed to determine the atomic composition of the first monolayer above the LSAT substrate (see Fig. S1 and Fig. S2). The Mn intensity measured at the first $Mn_2N$ layer at the interface was significantly lower as compared to $Mn_2N$ layers far from the interface. This difference in Mn-intensity, together with the observed alternating pattern of bright and dark spots in the HAADF-STEM image, points to a lower relative Mn concentration in every other atomic position (dark contrast spots) along the [100] direction in the interfacial monolayer. To quantify the Mn concentration at the interfacial monolayer HAADF-STEM image simulations, by changing the Mn occupancy (Fig. S3), was performed using the xHREM$^{TM}$ software (HREM research inc., Japan). The simulations are compatible with an approximately 80% Mn deficiency in the atomic positions corresponding to a darker contrast in the HAADF-STEM data. Thus, combining the simulations and structural and chemical analyses, the transition from the LSAT substrate to the $Mn_3GaN$ film is mediated by a sharp interfacial $Mn_xN$ monolayer with $x \sim 1.2$.

To unequivocally determine the atomic structure of the $Mn_xN$ interfacial monolayer additional STEM and EDS analyses along the [110] zone axis (Fig. S4) were performed. Fig. 4 shows a schematic of the proposed $Mn_3GaN$/LSAT interface based on analyses along the [100] and [110] zone axes. This model is also consistent with STEM analyses performed in epitaxial $Mn_3GaN$ films grown on (001)-oriented $SrTiO_3$ (see Supplementary Information). The fact that the projection of the $Mn_xN$ monolayer along the [100] and [010] directions are indistinguishable, the ordering of Mn and N atoms should constitute a 2-dimensional periodic structure with $C_4$ rotational symmetry. Considering $x = 1$ for simplicity, the ideal MnN monolayer would be arranged as depicted in Fig. 4B and 4C, with the N atoms located above (Al/Ta) atoms of LSAT and the Mn atoms over the interstice of the $(Al/Ta)O_2$ layer of LSAT. The illustration shows that the ideal

MnN interfacial monolayer has an analogous structure as a perovskite *A*O layer, with *A* being Mn and N being O. Moving away from the interface, a MnGa puckered layer is observed occurring on top of the MnN interfacial layer (Fig. 3B), with the Mn cations displaced towards the interface. A gradual decrease of the interplanar distances along the [001] direction within the first five layers of above the interface is also apparent. Above the fifth layer the interplanar distance reaches the bulk value.

To study the stability of the interfacial model derived from the atomic resolutions experiments first principles calculations were performed. Due to the complex crystal structure of LSAT, $AlO_2$-terminated $LaAlO_3$ was employed to mimic $BO_2$-terminated LSAT. $Mn_3GaN/LaAlO_3$ with two different interfaces, $MnN/AlO_2$ and $Mn_2N/AlO_2$ were simulated as shown in Fig. 5A. Specifically, their formation energies $\Delta E$ were calculated to test for stability. As shown in Table 1, the calculation results indicate that both interfaces have negative $\Delta E$, which implies that both are energetically stable. However, $\Delta E$ = - 2.265 eV for the $Mn_2N$ interface is appreciably lower than that of the observed MnN interface, $\Delta E$ = - 0.058 eV. The lower $\Delta E$ for the $Mn_2N$ interface can be understood from the chemical bonding at the interface. As shown in Fig. 5B and Fig. 5C, the charge density between Mn and O (or N) at the MnN interface is significantly smaller than that of the $Mn_2N$ interface, corresponding to a stronger Mn-O and Mn-N bonding at a $Mn_2N$ interface, thus resulting in a more cohesive and energetically stable interface.

The apparent discrepancy between the interfacial models derived from the experimental and theoretical studies can be solved by taking into account the on-set of $Mn_3GaN$ growth in the presence of an energy barrier preventing the system from relaxing from the local to the global energy minimum. To explore this hypothesis, the formation energies for $Mn/LaAlO_3$ was calculated using two different Mn configurations, Mn(1) and Mn(2) as shown in Fig. S10. Mn(1) and Mn(2) correspond to the positions of Mn in the MnN and $Mn_2N$ interfaces, respectively, as described in Fig. 5D. Interestingly, the $Mn/LaAlO_3$ supercell with the Mn atom located in the Mn(1) site had a lower energy than that of the Mn atom at the Mn(2) position (Table 1). An analogous behavior was observed by calculations using the non-polar $SrTiO_3$ surface, which indicates that the Mn(1) site is the most energetically favorable position for Mn on both polar and non-polar $ABO_3$ perovskite surfaces. While Mn(1) is surrounded by four $O^{2-}$ anions, in the vicinity of Mn(2) there is one $O^{2-}$ and two B cations. The strong local Coulomb repulsion between Mn(2) and the B cations accounts for the higher formation energy of the $Mn(2)/ABO_3$ supercells. Additionally, the more positive the B cation, the higher energy of the Mn(2) site. As is evidenced in Table 1, showing that the calculated energy difference between $Mn(1)/SrTiO_3$ and $Mn(2)/SrTiO_3$ (0.452 eV) is larger than that of $Mn(1)/LaAlO_3$ and $Mn(2)/LaAlO_3$ (0.155 eV), mainly due to $Ti^{4+}$ being more positive than $Al^{3+}$. Thus by conjecture, the Mn(2) site in the Mn(2)/LSAT system will hence be more unstable due to the $Ta^{5+}$ cations in the LSAT terminating layer.

That is, during the initial growth of $Mn_3GaN$, Mn ions arriving at the B-terminated $ABO_3$ layer sit on the Mn(1) positions and then coordinate with N, forming a $Mn_xN$ monolayer as determined by the STEM studies. This interfacial monolayer works as a structural bridge between the *AB*$O_3$ substrate and $Mn_3GaN$ film and establishes heteroepitaxy between the two non-isostructural materials with different chemical composition and bonding. Moreover, the experimentally observed puckered GaMn layer can be related to the strong out-of-plane Ga-Mn bonding due to the strong charge density overlap between Mn in the first interfacial layer and Ga in the layer 2 (Fig. 5B) as shown in Fig. S11.

## Discussion

The realization of an atomically sharp bridging structure allowing an epitaxial interface structure and bonding between nitride antiperovskites and oxide perovskites manifests a critical step in the development of a new class of epitaxial heterostructures based on materials with dissimilar crystallochemical properties. The ability to engineer such novel heterointerfaces from chemically divergent constituents brings a new dimension to the mature field of complex oxides, and provides a playground for the manipulation of the interfacial physical properties, as well as the establishment of new states of matter. In particular, Mn-based nitride antiperovskites with non-collinear $\Gamma^{5g}$ triangular antiferromagnetic structures are ideal systems to interface with piezoelectric or ferroelectric oxide compounds to induce piezomagnetic or magnetoelectric effects in the antiperovskite, as recently proposed theoretically (16-18). Additionally, materials showing geometrically frustrated antiferromagnetic spin structures like $Mn_3GaN$ are the source of intriguing physical behavior including large anomalous Hall and Nernst effects, large magnetoresistance, spin transfer torque, and spin Hall effect (21-24). Given the potential of these materials for antiferromagnetic spintronics (25), the rational design of epitaxial heterostructures of Mn-based nitride antiperovskites and *AB*$O_3$ perovskites is of great importance for property tuning and functional device design. We expect our study to trigger the investigation and development of functional antiperovskite/perovskite heterostructures, opening a new and exciting avenue for materials design.

## Materials and methods

**Sample growth and x-ray characterization**. Thin film heterostructures were grown by DC reactive planar magnetron sputtering using a $Mn_3Ga$ stoichiometric target (99.9% purity) at 50 W. The films were deposited at substrate temperature of 550 °C in an Ar (50 sccm)/$N_2$ (5.2 sccm) atmosphere of 9.5 mTorr. The sample to target distance was fixed to 4 inches. Prior to deposition, the vacuum chamber was evacuated until a base pressure of $10^{-7}$ Torr was achieved. X-ray characterization of the samples was performed at room temperature by using four-circle X-ray diffractometer equipped with Cu–K$\alpha_1$ radiation.

**HAADF-STEM imaging and atomic-resolution EDS and EELS.** Due to the delicate bonding between antiperovskite nitride and perovskite oxide, samples for STEM observation should be carefully prepared. Focus ion beam (FIB) sampling or prolonged exposure to ion-milling caused the interface to collapse and the damaged area looks dark with a few nanometers thickness along the interface. Therefore, samples were prepared via the conventional way. Samples were mechanically grinded to a thickness of less than 50 μm (EM TXP, Leica, Germany), dimpled to a thickness of ~ 5 μm (Dimple Grinder II, Gatan, , USA), and thinned for electron transparency by Ar ion-beam milling with $LN_2$-cooling stage (Precision Ion Polishing System II, Gatan, USA). HAADF (high angle annular dark field)-STEM images were taken in a scanning transmission electron microscope (JEM-2100F, JEOL) at 120 kV with a spherical aberration corrector (CEOS GmbH). The optimum size of the electron probe was ~ 1.2 Å. The collection semi angles of the HAADF detector were adjusted from 70 to 240 mrad. The obtained raw images were band-pass filtered to reduce background noise (HREM Filters Pro, HREM research, Japan). To identify the interfacial chemistry, energy loss spectra were obtained in JEM-2100F (JEOL) at 120 kV using an EEL spectrometer (GIF Quantum ER, Gatan, USA). Because Ga, Sr, Ta, and Al species are not detectable via EELS, the further chemistry at the interface was understood via the atomic-level energy dispersive spectroscopy (EDS) with a 100 $mm^2$ detector (X-max$^N$, Oxford, United Kingdom).

**Computational details**. First-principles calculations were performed with the projector augmented-wave (PAW) method (26) implemented in VASP code[27] using unconstrained non-collinear magnetic structures (28,29). The exchange and correlation effects were treated within the generalized gradient approximation (GGA) (30). We used the plane-wave cut-off energy of 550 eV and a 16 × 16 × 16 and 12 × 12 × 1 *k*-point meshes in the irreducible Brillouin zone for bulk and interface structures, respectively. Two supercells of $Mn_3GaN/LaAlO_3$ (with the formula $Mn_{12}Ga_4N_5La_4Al_5O_{14}$ for MnN phase and $Mn_{14}Ga_4N_5La_4Al_5O_{14}$ for $Mn_2N$ phase) were used to simulate the interfacial structure (Fig. 5A). Since previous reports showed that magnetism strongly influences the calculated lattice constant in $Mn_3GaN$ (16), when optimizing the lattice structure, we assumed a non-collinear magnetic order in bulk $Mn_3GaN$, while the interfacial MnN layer was set to be antiferromagnetically aligned to the neighboring GaMn layer. The in-plane lattice constant of the interface supercell was constrained to the calculated lattice constant of bulk cubic $Mn_3GaN$ ($a$ = 3.867 Å). The internal coordinates and the $c$ lattice constant were relaxed until the force on each atom was less than 0.001 eV/Å. When evaluating the stability of $Mn/ABO_3$, we used the symmetric supercells (with the formula $Mn_2A_4B_5O_{14}$) made by $ABO_3$ slab, Mn monolayers and a vacuum layer over 15 Å, as shown in Fig. S10.

The formation energies were evaluated as follows (31):

$$\Delta E_{MnN_interface} = (E_{supercell} - 4E_{Mn_3GaN} - 4E_{LaAlO_3} - E_{Al} - E_N - 2E_O)/2,$$
$$\Delta E_{Mn_2N_interface} = (E_{supercell} - 4E_{Mn_3GaN} - 4E_{LaAlO_3} - E_{Al} - E_N - 2E_O - 2E_{Mn})/2,$$
$$\Delta E_{Mn/LaAlO_3} = (E_{supercell} - 4E_{LaAlO_3} - 2E_{Mn} - E_{Al} - 2E_O)/2,$$

where the $E_{Mn_3GaN}$, $E_{LaAlO_3}$, $E_B$, and $E_{Mn}$ are the total energies of the related bulk material, $E_N$ and $E_O$ are the half of the total energies of the related molecule.

## Supplementary Materials

Fig. S1. [100]-projected HAADF-STEM images of the $Mn_3GaN$/LSAT interface and corresponding recorded EELS areal density maps.

Fig. S2. [100]-projected HAADF-STEM images of the $Mn_3GaN$/LSAT interface and corresponding recorded EDS.

Fig. S3. [100]-projected HAADF-STEM image of the $Mn_3GaN$/LSAT interface showing the atomic rows where the intensity profile.

Fig. S4. [110]-projected HAADF-STEM image of the $Mn_3GaN$/LSAT and corresponding recorded EELS areal density maps.

Fig. S5. X-ray diffraction spectroscopy data for $Mn_3GaN/SrTiO_3$ samples.

Fig. S6. [100]-projected HAADF-STEM images of the $Mn_3GaN/SrTiO_3$ interface and corresponding recorded EELS areal density maps.

Fig. S7. [100]-projected HAADF-STEM images of the $Mn_3GaN/SrTiO_3$ interface and corresponding recorded EDS data.

Fig. S8. [100]-projected HAADF-STEM image and model around the $Mn_3GaN/SrTiO_3$ interface.

Fig. S9. [110]-projected HAADF-STEM image and model around the $Mn_3GaN/SrTiO_3$ interface.

Fig. S10. $Mn/ABO_3$ supercells used for the theoretical calculations.

Fig. S11. Computational calculated model of the puckered GaMn layer.

## Acknowledgement

**Funding:** This work was supported by the National Science Foundation under DMREF Grant No. DMR-1629270. Transport measurement at the University of Wisconsin–Madison was supported by the US Department of Energy (DOE), Office of Science, Office of Basic Energy Sciences (BES), under award number DE-FG02-06ER46327. **Author contributions:** C.B.E. conceived the project and directed the research. C.B.E., M.S.R., E.Y.T., S.H.O. and L.J.B. supervised the experiments. C.X.Q. and T.N. fabricated and characterized thin film samples. N.C. and M.S.R. carried out electrical transport measurements. K.S., S.Y.C., L.X. and X.Q.P. carried out scanning transmission electron microscopy. D.F.S., T.R.P. and E.Y.T. performed theoretical calculations. All authors wrote and commented on the paper. C.B.E. directed the overall research. **Competing interests:** The authors declare that they have no competing interests. **Data and materials availability:** All data needed to evaluate the conclusions in the paper are present in the paper and/or the Supplementary Materials. Additional data related to this paper may be requested from the authors.

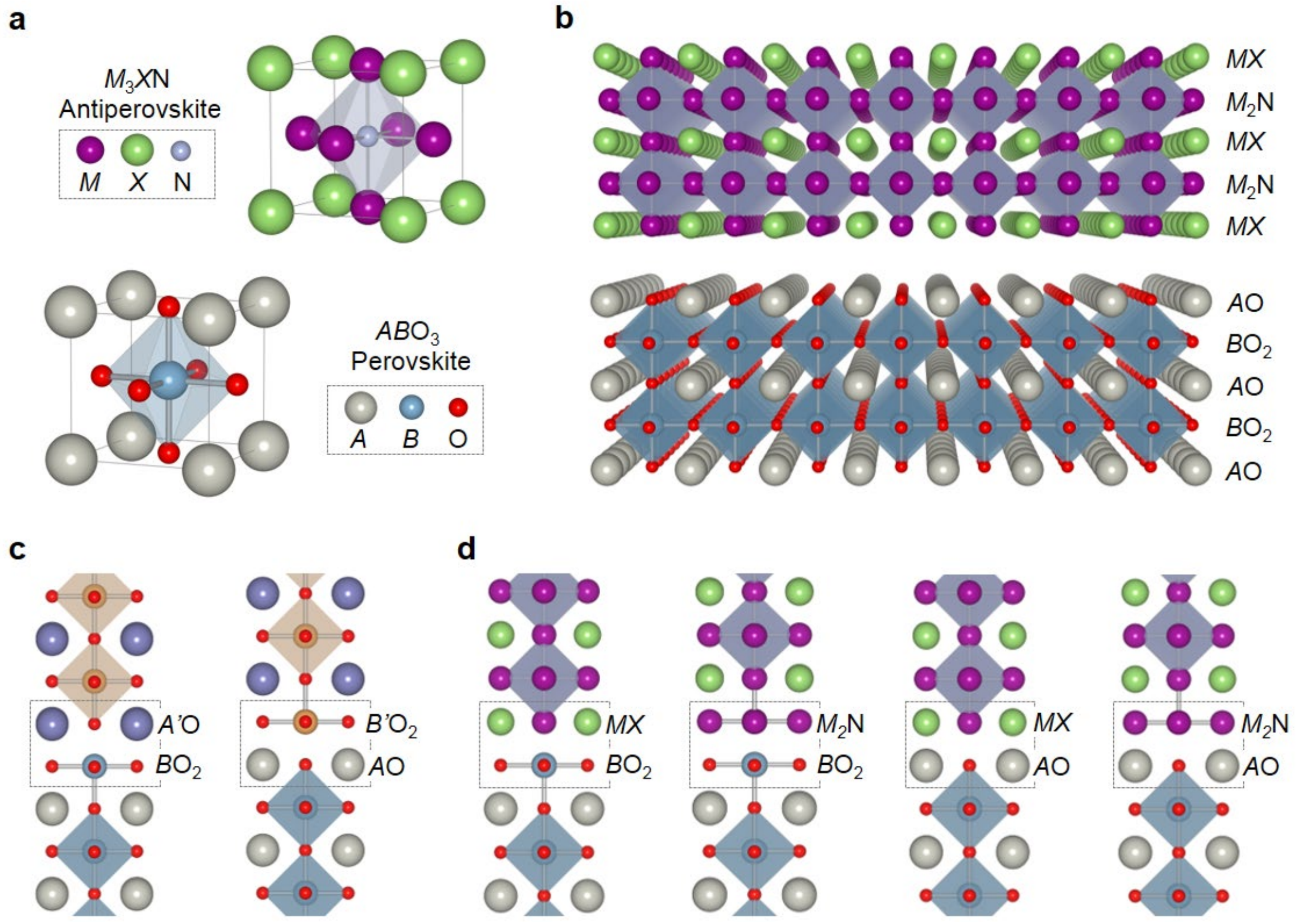

**Fig. 1. Schematic representation of the crystal structures of *M*$_3$XN nitride-antiperovskite and *AB*O$_3$ oxide-perovskite compounds and their interfaces**. (**A**) $M_3X$N and $AB$O$_3$ ideal unit cells showing their geometrically analogous crystal structures and reversed anion (N, and O) and cation (*M*, and *B*) positions in the unit cell. (**B**) $M_3X$N and $AB$O$_3$ slabs represented as a stacking of alternating *A*O and $B$O$_2$, and $M_2$N and *MX* planes, respectively. (**C**) Representation of the two proven atomically sharp interfacial configurations (*A*'O:$B$O$_2$ and *B*'O$_2$:*A*O) between two different oxide perovskite compounds $AB$O$_3$ and *A'B'*O$_3$. (**D**) Representation of the four-possible atomically abrupt interfacial configurations (*MX*:$B$O$_2$, $M_2$N:$B$O$_2$, *MX*:*A*O, and $M_2$N:*A*O) between $AB$O$_3$ and $M_3X$N compounds depending on the $AB$O$_3$ termination layer.

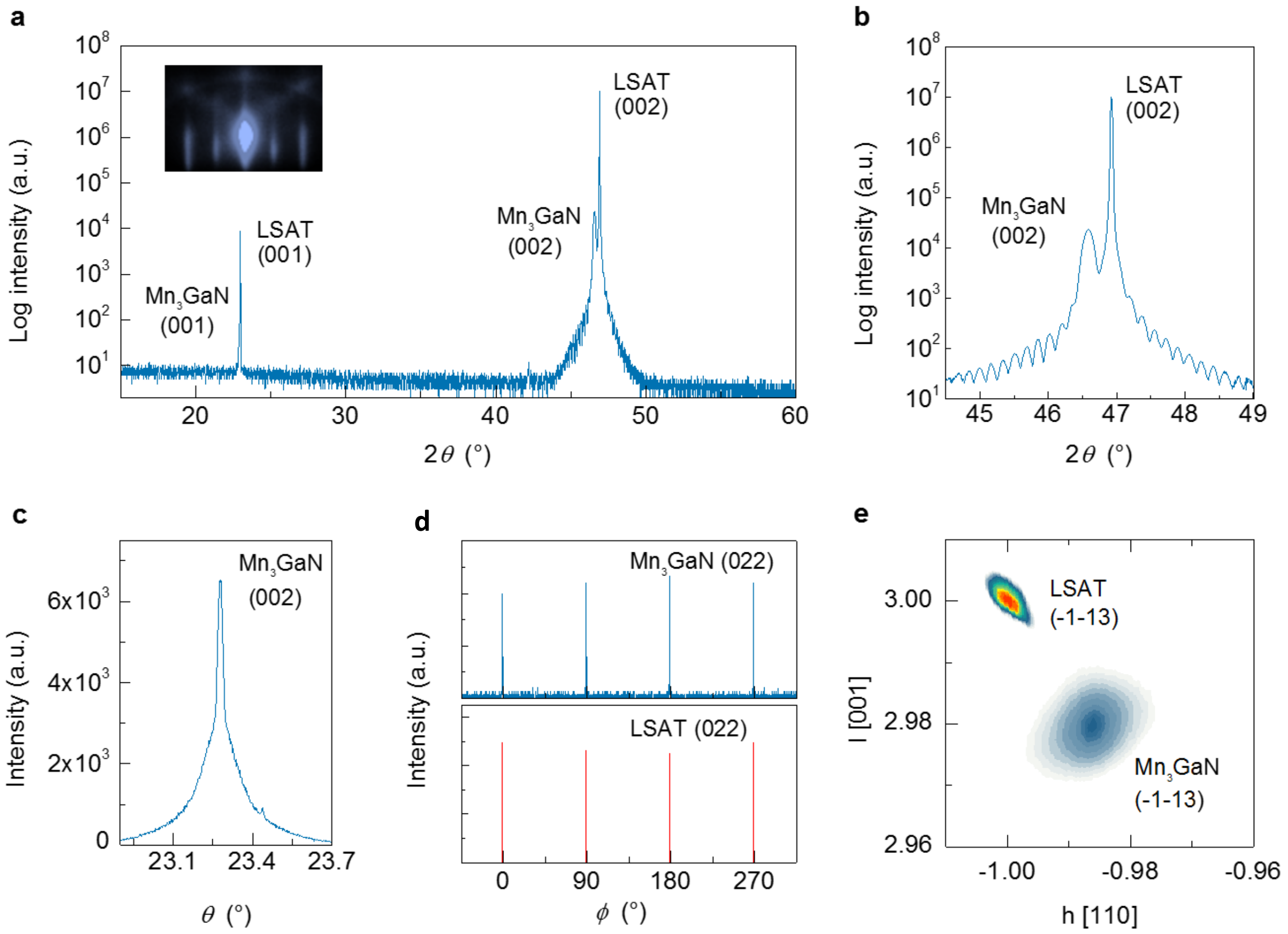

**Fig. 2. X-ray diffraction structural characterization of a 60 nm thick $Mn_3GaN$ grown on a (001)-oriented LSAT substrate.** (**A**) Wide angle θ-2θ spectrum only shows the (00*l*) reflections of LSAT substrate and $Mn_3GaN$ film, demonstrating the film is (001)-oriented and single phase. Inset shows registered RHEED pattern of the specular diffraction spot after growth. (**B**) Short range θ-2θ scan around the (002) diffraction peak of the $Mn_3GaN$ film showing Kiessig fringes, indicating pristine interfaces and high crystalline quality of the film. (**C**) Rocking curve of the (002) $Mn_3GaN$ peak. (**D**) 360° ϕ-scans around the $Mn_3GaN$ and LSAT (022) peaks demonstrates cube on cube epitaxial relationship. (**E**) RSM around the LSAT (-113) reciprocal lattice point shows the $Mn_3GaN$ is strain relaxed.

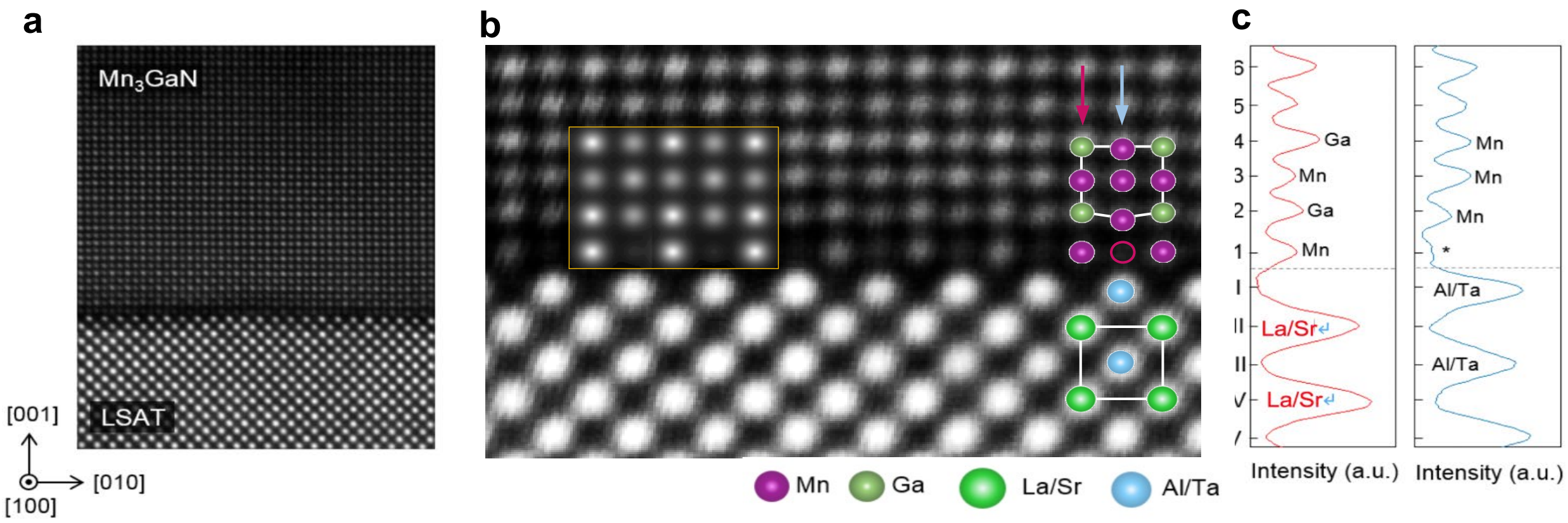

**Fig. 3. HAADF-STEM characterization of the $Mn_3GaN$/LSAT heterointerface.** (**A**) HAADF-STEM imaging of the $Mn_3GaN$/LSAT heterostructure taken along the [100] zone axes of LSAT. (**B**) Magnified HAADF-STEM imaging overlaid with the cation positions and simulated image of the interface (yellow square). Orange lines are a guide to the eyes showing buckling of the Mn and Ga atoms at the second row. (**C**) Integrated HAADF-STEM intensity line profile along two adjacent atomic columns (out-of-plane direction, represented by arrows in (**B**)). Ordinate y-axis shows the layer's number, denoted by roman numerals for LSAT and arabic numerals for $Mn_3GaN$. Since the HAADF-STEM intensity is proportional to $Z^2$ (Z: atomic number), Ga atoms show higher intensity than Mn atoms. The * symbol indicates Mn-deficiency.

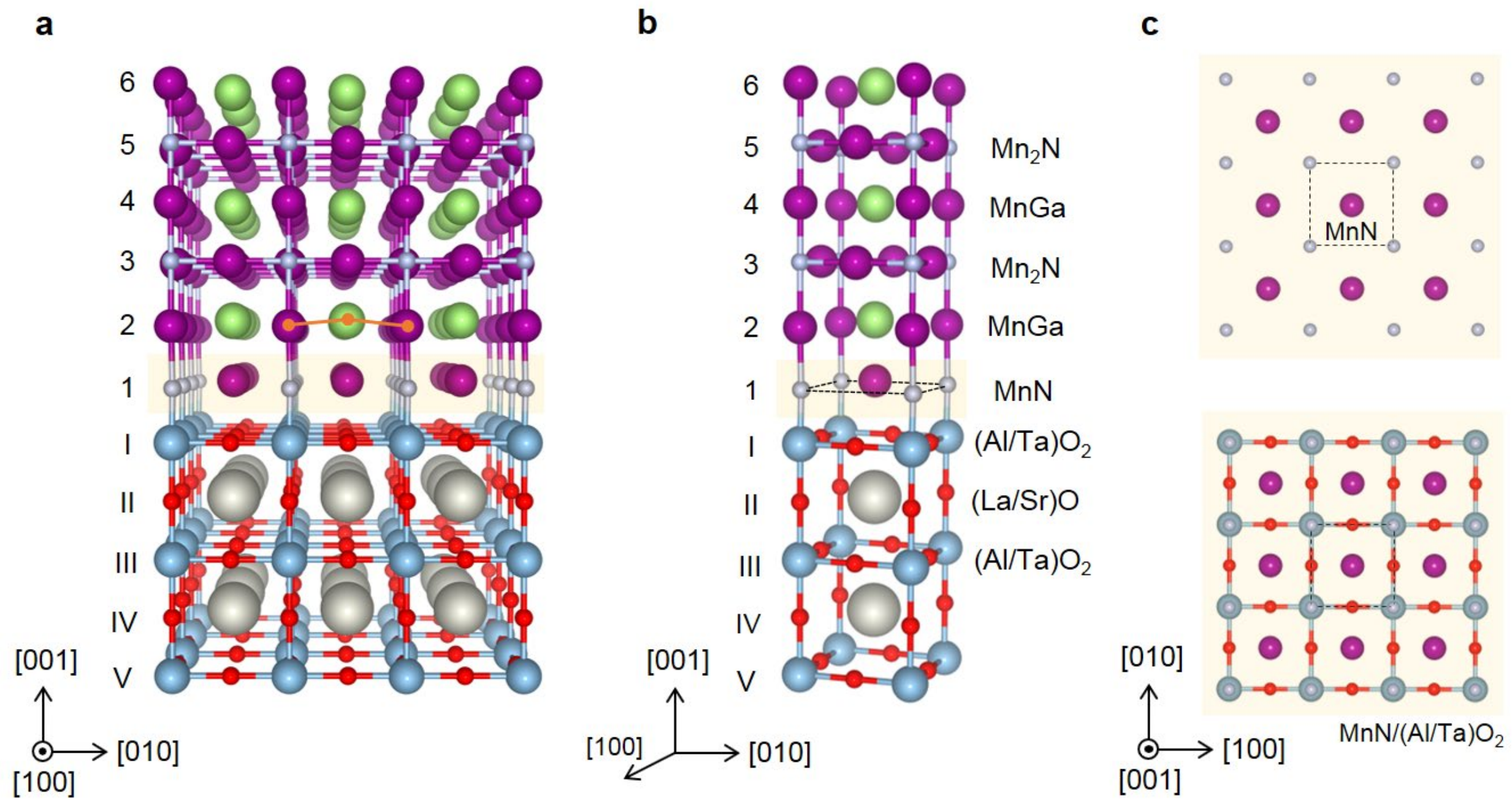

**Fig. 4. Illustration of the $Mn_3GaN$/LSAT heterointerface based on our experimental results.** (**A**) Schematic [100] perspective view of the $Mn_3GaN$/LSAT heterointerface. Orange line in layer 2 is a guide to the eyes showing buckling of the Mn and Ga atoms. (**B**) Representation of the $Mn_3GaN$/LSAT heterointerface as a stacking of atomic unit cell planes. (**C**) [001] projections of the MnN interfacial layer (top image) and MnN layer overlaid with the (Al/Ta)$O_2$ LSAT termination layer (bottom image). Dashed square represents the interfacial MnN unit cell.

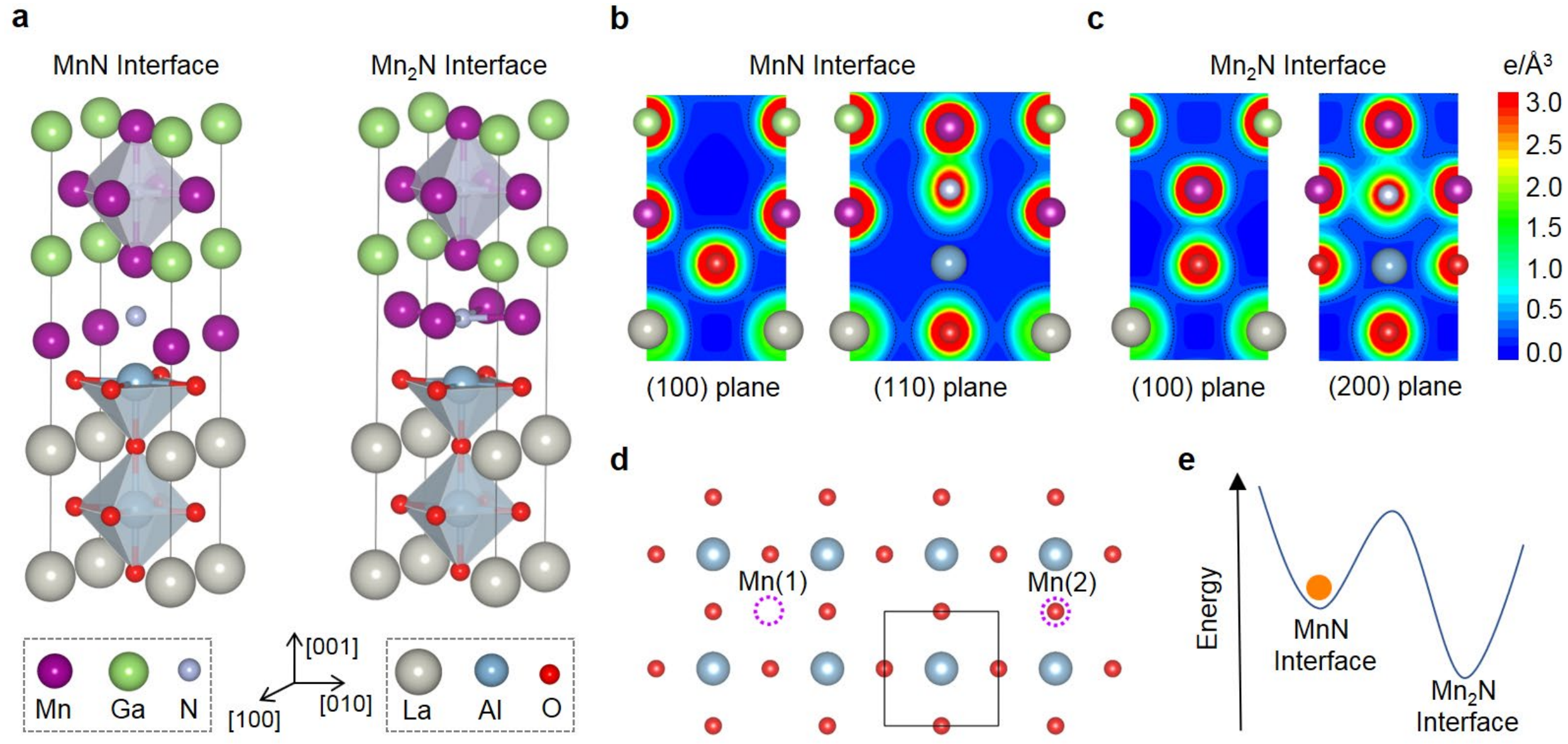

**Fig. 5. Theoretical calculations for different interfacial configurations.** (**A**) Sections of the relaxed $Mn_3GaN/LaAlO_3$ supercell with the $MnN/AlO_2$ interface and **b,** the $Mn_2N/AlO_2$ interface. (**B**) Charge density plots around the MnN interface in the (100) and (110) planes. (**C**) Charge density plots around the $Mn_2N$ interface in the (100) and (200) planes. (**D**) Illustration of the two possible deposited positions of Mn atoms Mn(1) and Mn(2) (purple dashed circles) onto the $AlO_2$ plane. (**E**) Schematic diagram of energies of MnN interface and $Mn_2N$ interface, showing that the MnN interface is in a local energy minimum.

**Table 1. Formation energy for different interfacial configuration.** Calculated formation energies ΔE of $Mn_3GaN/LaAlO_3$ for two interfacial configurations: MnN and $Mn_2N$. Calculated ΔE for Mn deposited in the Mn(1) and Mn(2) positions onto $BO_2$ terminated $LaAlO_3$ and $SrTiO_3$ .

| | $Mn_3GaN/LaAlO_3$ (MnN interface) | $Mn_3GaN/LaAlO_3$ ($Mn_2N$ interface) | Mn(1)/$LaAlO_3$ | Mn(2)/$LaAlO_3$ | Mn(1)/$SrTiO_3$ | Mn(2)/$SrTiO_3$ |
|---|---|---|---|---|---|---|
| $\Delta E$ (eV /interface) | −0.058 | −2.265 | −0.980 | −0.825 | −1.698 | −1.246 |

**Supporting Information for:**

# Epitaxial Antiperovskite/Perovskite Heterostructures for Materials Design

Camilo X. Quintela,[1] Kyung Song,[2] Ding-Fu Shao,[3] Lin Xie,[4] Tianxiang Nan,[1] Tula R. Paudel,[3] Neil Campbell,[5] Xiaoqing Pan,[7] Thomas Tybell,[8] Mark S. Rzchowski,[5] Evgeny Y. Tsymbal,[3] Si-Young Choi,[6*] Chang-Beom Eom[1*]

[1] Department of Materials Science and Engineering, University of Wisconsin-Madison, Madison, Wisconsin 53706, USA. [2] Department of Materials Modeling and Characterization, KIMS, Changwon 51508, South Korea. [3] Department of Physics and Astronomy & Nebraska Center for Materials and Nanoscience, University of Nebraska, Lincoln, Nebraska 68588, USA. [4] National Laboratory of Solid State Microstructures and College of Engineering and Applied Sciences, Nanjing University, Nanjing, Jiangsu 210093, People's Republic of China. [5] Department of Physics, University of Wisconsin-Madison, Madison, Wisconsin 53706, USA. [6] Department of Materials Science and Engineering, POSTECH, Pohang 37673, South Korea. [7] Department of Chemical Engineering and Materials Science, University of California-Irvine, Irvine, CA 92697, USA. [8] Department of Electronic Systems, Norwegian University of Science and Technology, Trondheim 7491, Norway;

*Corresponding author. Email: eom@engr.wisc.edu, youngchoi@postech.ac.kr

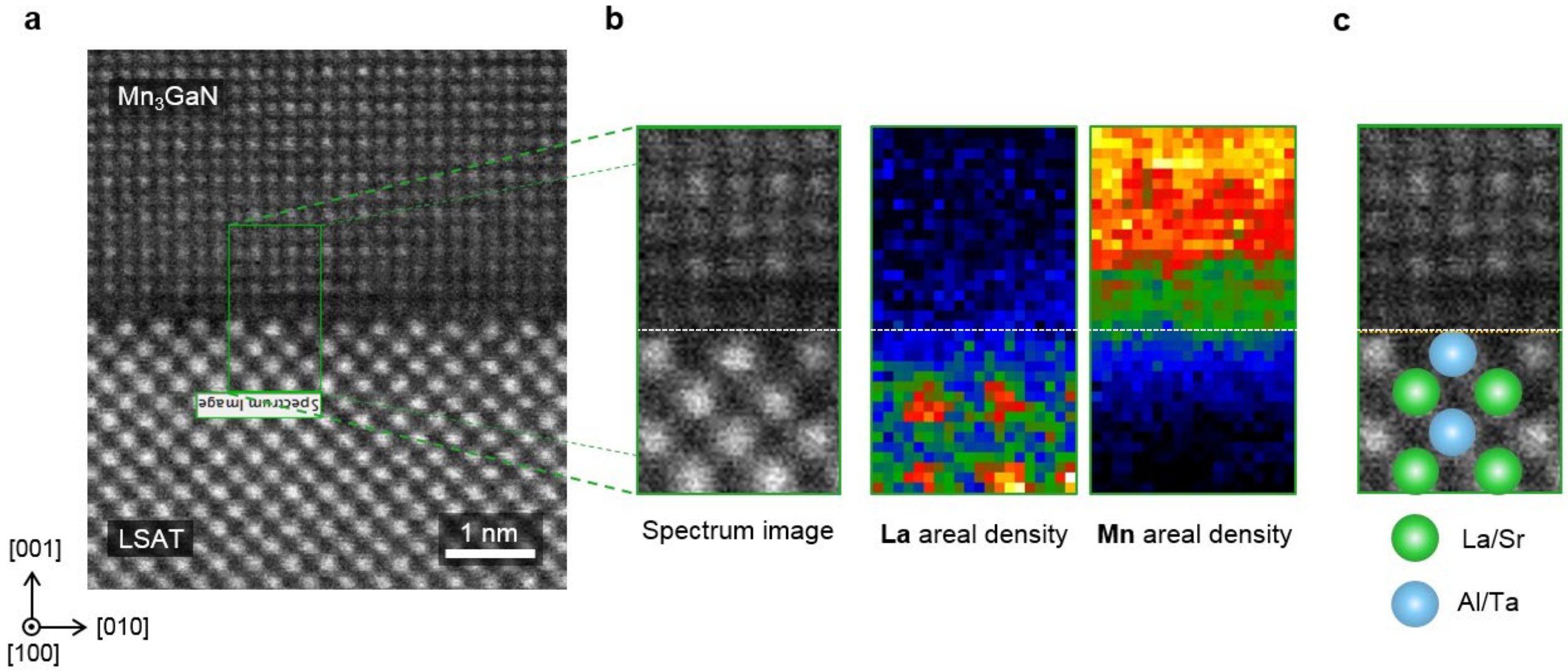

Fig S1. (**A**) [100]-projected HAADF-STEM image of the $Mn_3GaN$/LSAT interface with the area from which EELS mapping was carried out, highlighted with a green line rectangle. (**B**) EELS areal density maps for La and Mn carried out in the area labeled "Spectrum image". (**C**) HAADF-STEM "Spectrum image" overlaid with the proposed LSAT termination based on EELS analysis. The yellow dashed line marks the interface between $Mn_3GaN$ and LSAT.

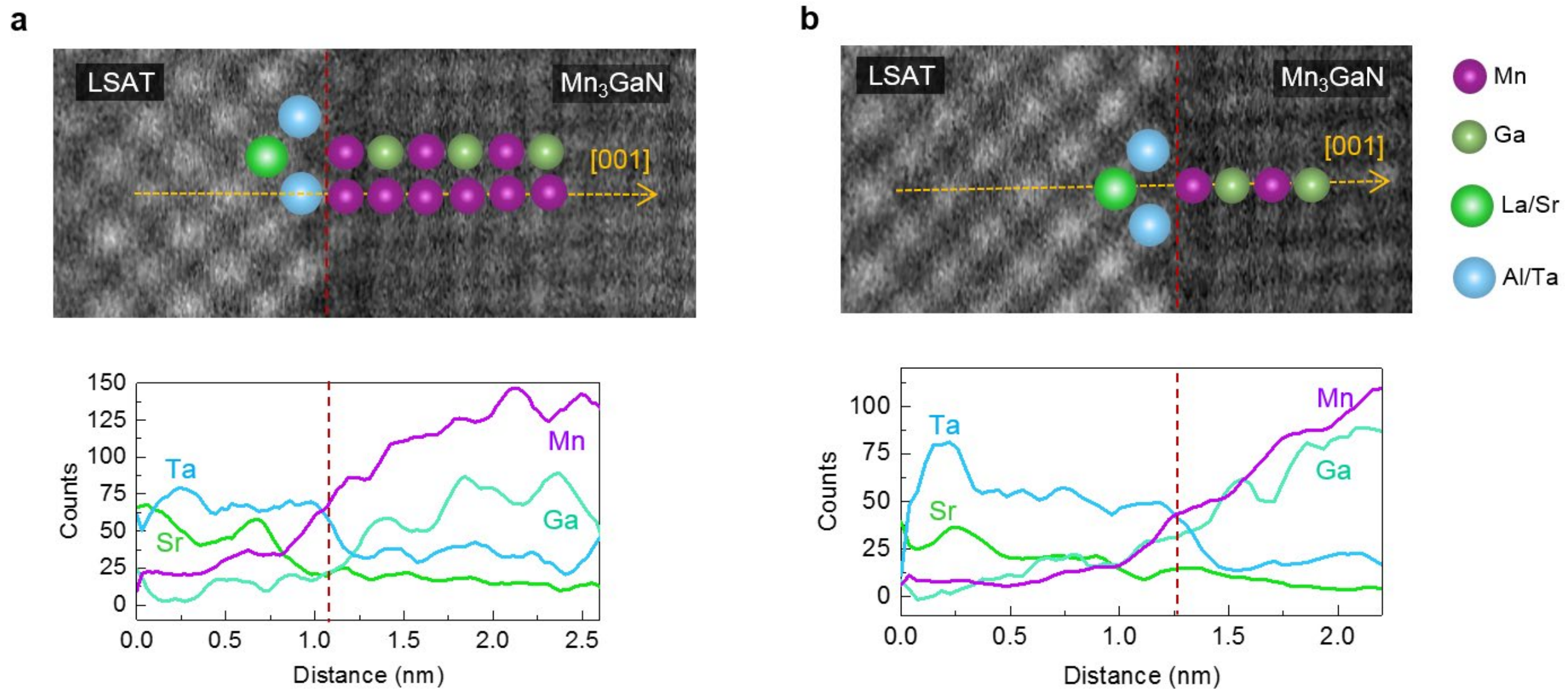

Fig. S2. (**A**) and (**B**) [100]-projected HAADF-STEM images of the $Mn_3GaN$/LSAT interface and (below each image) corresponding recorded EDS data along the atomic rows represented by yellow arrows in the HAADF-STEM image. EDS analysis show a dominant Mn signal at the interface. Overlaid on the HAADF-STEM images is the proposed atomic configuration at the interface based on EELS and EDS analyses.

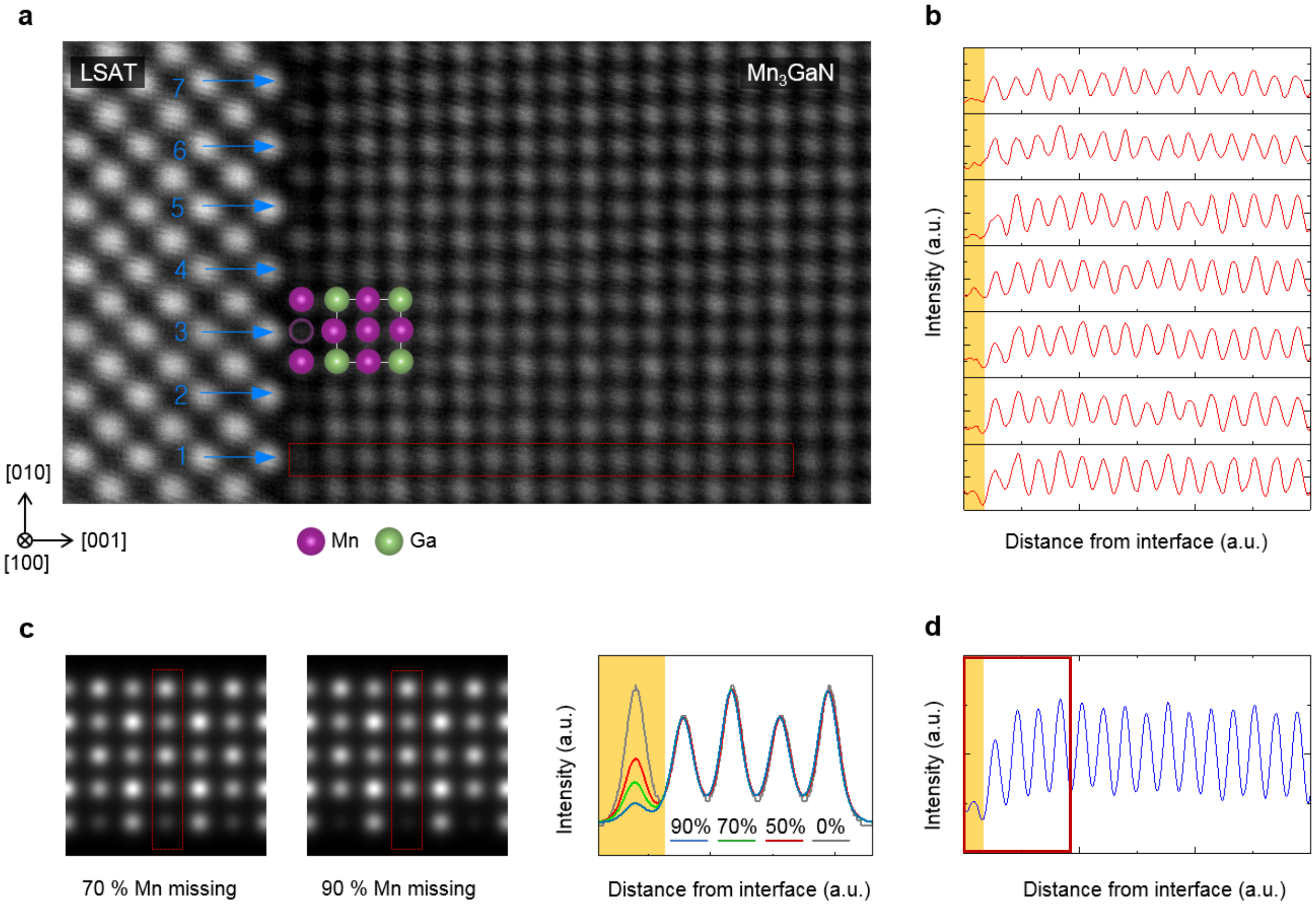

Fig. S3. (**A**) [100]-projected HAADF-STEM image of the $Mn_3GaN$/LSAT interface showing the atomic rows where the intensity profile was measured, as shown in graphs (**B**) (**C**) Simulated image for 70% and 90% Mn atoms missing at the interface and simulated intensity profile for 90%, 70%, 50%, and 0% Mn atoms missing. (**D**) Average intensity profile calculated from experimental data shown in b. By comparing experimental and simulated intensities, the estimated Mn content on every other atomic position at the interfacial monolayer is about 10 %.

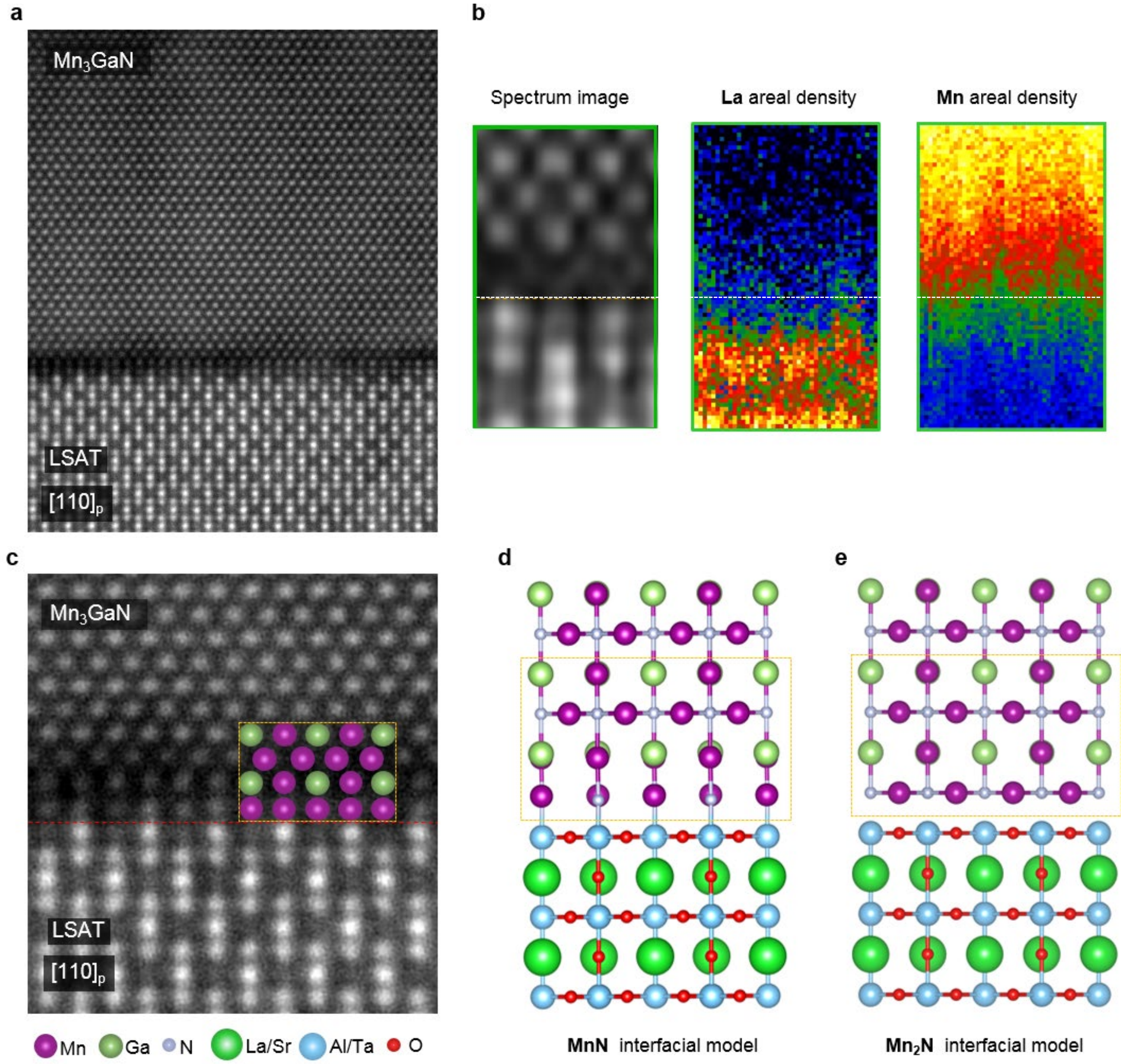

Fig. S4. (**A**) [110]-projected HAADF-STEM image of the $Mn_3GaN$/LSAT (**B**) EELS areal density maps for La and Mn carried out in the area named as "Spectrum image" (**C**) [110]-projected HAADF-STEM image of the $Mn_3GaN$/LSAT overlaid with the proposed atomic configuration for $Mn_3GaN$ at the interface based on EDS and EELS analyses. (**D**) [110]-projected atomic model assuming that the first monolayer is "MnN". This model matches the atomic configuration derived from experimental data. (**E**) [110]-projected atomic model assuming that the first monolayer is $Mn_2N$. This model does not match the atomic configuration derived from experimental data.

## 1. Supplementary information and figures for $Mn_3GaN$ films grown on (001)-oriented $SrTiO_3$

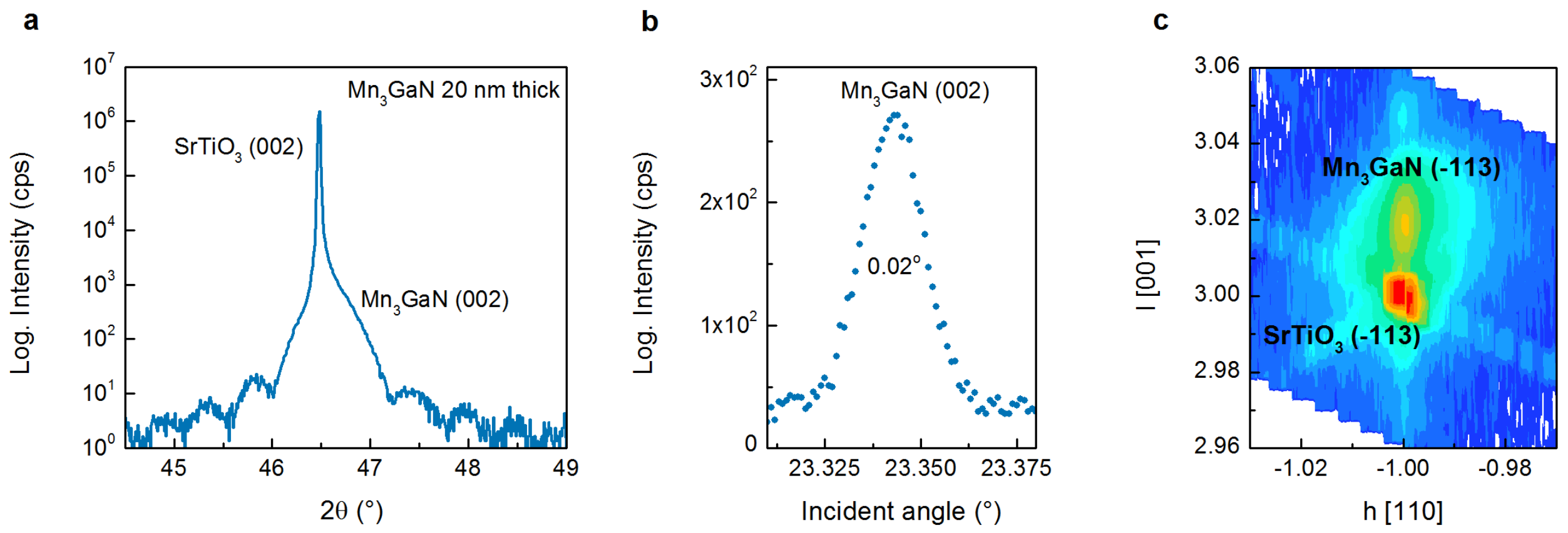

Fig. S5. (**A**) Out-of-plane x-ray diffraction scan around the (002) diffraction peak of the $Mn_3GaN$ film. (**B**) Rocking curve measured around the (002) diffraction peak of $Mn_3GaN$. The FWHM value of 0.02° attest to the high crystalline quality of the film, (**C**) X-ray reciprocal space map measured around the $SrTiO_3$ (-113) reciprocal lattice point showing the coherent growth of $Mn_3GaN$ on $SrTiO_3$.

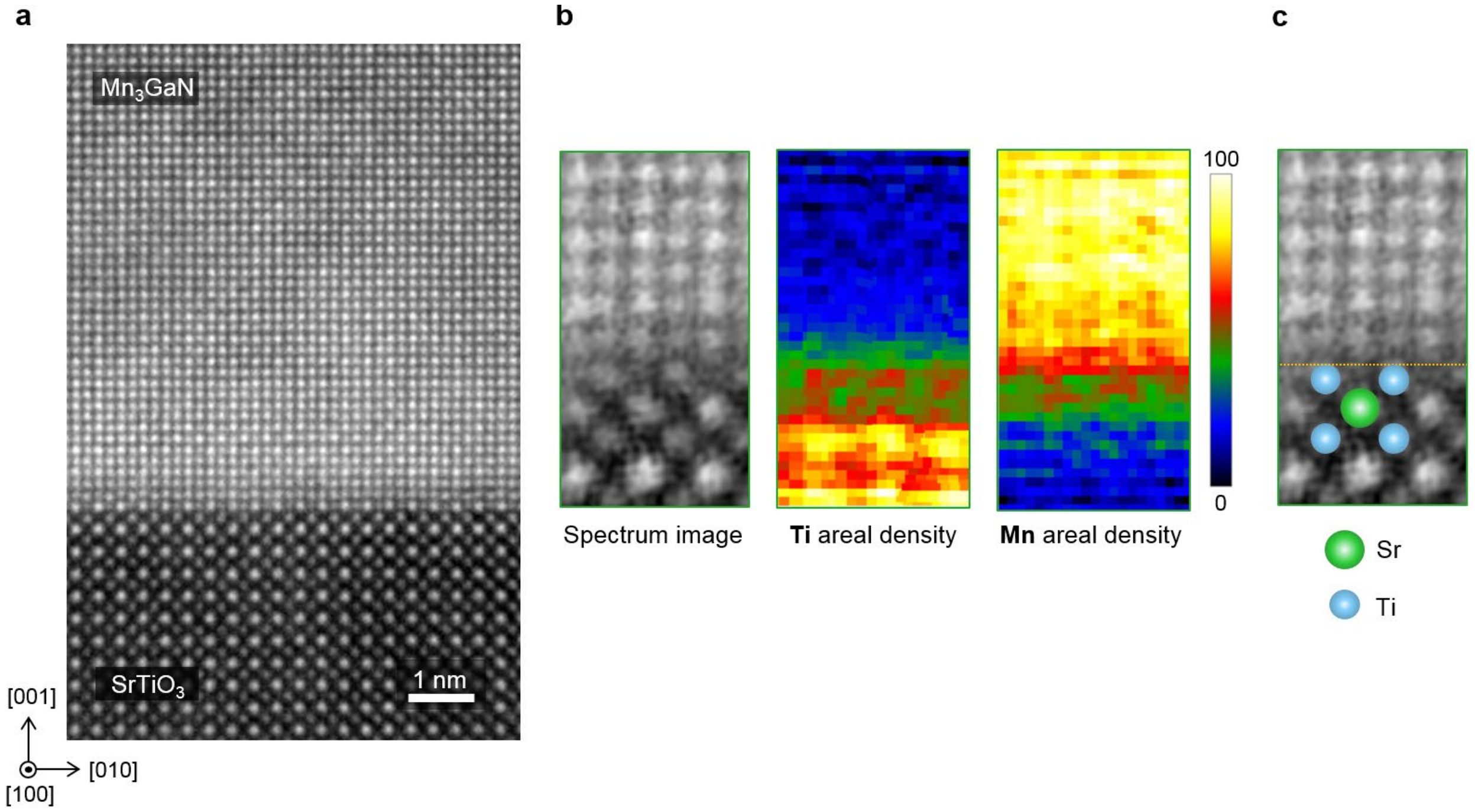

Fig. S6. (**A**) [100]-projected HAADF-STEM image of the $Mn_3GaN/SrTiO_3$ interface (**B**) EELS areal density maps for Ti and Mn carried out in the area named as "Spectrum image". (**C**) HAADF-STEM "Spectrum image" overlaid with the proposed $SrTiO_3$ termination based on EELS analysis. The yellow dashed line marks the interface between $Mn_3GaN$ and $SrTiO_3$.

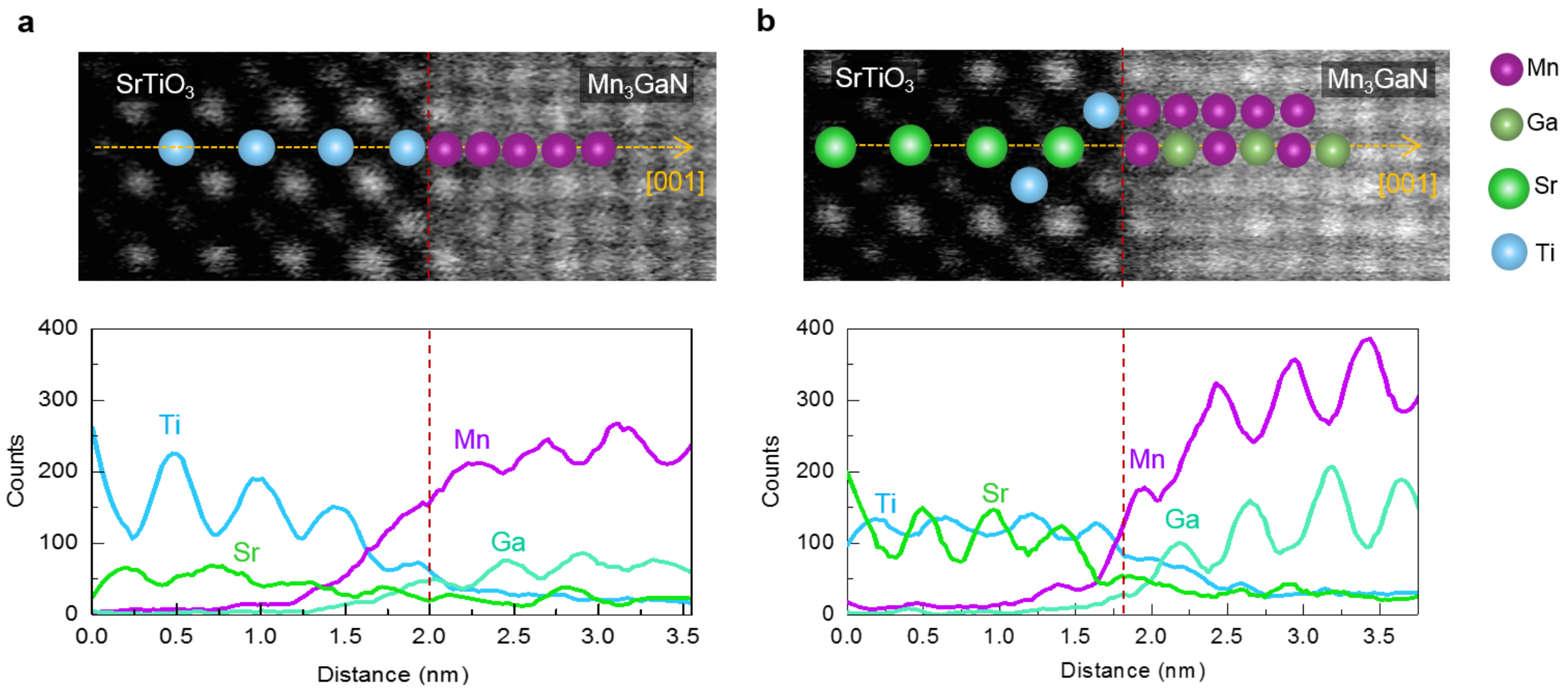

Fig. S7. (**A**) and (**B**) [100]-projected HAADF-STEM images of the $Mn_3GaN/SrTiO_3$ interface and (below each image) corresponding recorded EDS data along the atomic rows represented by yellow arrows in the HAADF-STEM image. EDS analysis show a dominant Mn signal at the interface. Overlaid on the HAADF-STEM images is the proposed atomic configuration at the interface based on EELS and EDS analyses.

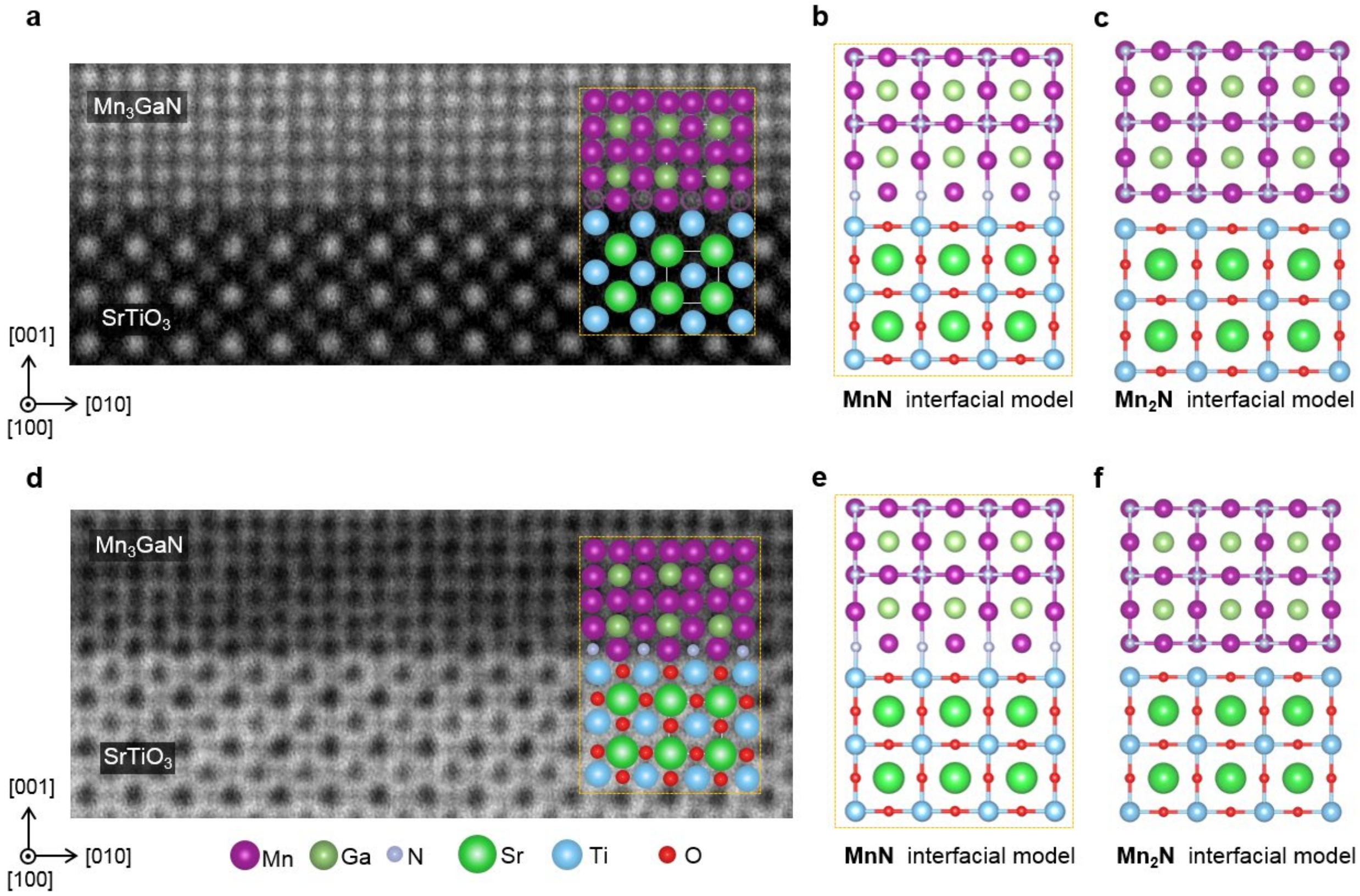

Fig. S8. (**A**) [100]-projected HAADF-STEM image around the $Mn_3GaN/SrTiO_3$ interface. Overlaid on the HAADF-STEM image is the proposed atomic configuration at the interface based on EELS and EDS analyses. (**B**) [100]-projection of the interfacial model considering a MnN interfacial layer. (**C**) [100]-projection of the interfacial model considering a $Mn_2N$ interfacial layer. (**D**) [100]-projected ABF-STEM image around the $Mn_3GaN/SrTiO_3$ interface. Overlaid on the ABF-STEM image is the proposed atomic configuration at the interface, which matches with the interfacial model considering a MnN interfacial layer as shown in (**E**) but disagrees with the $Mn_2N$ interfacial model as shown in (**F**).

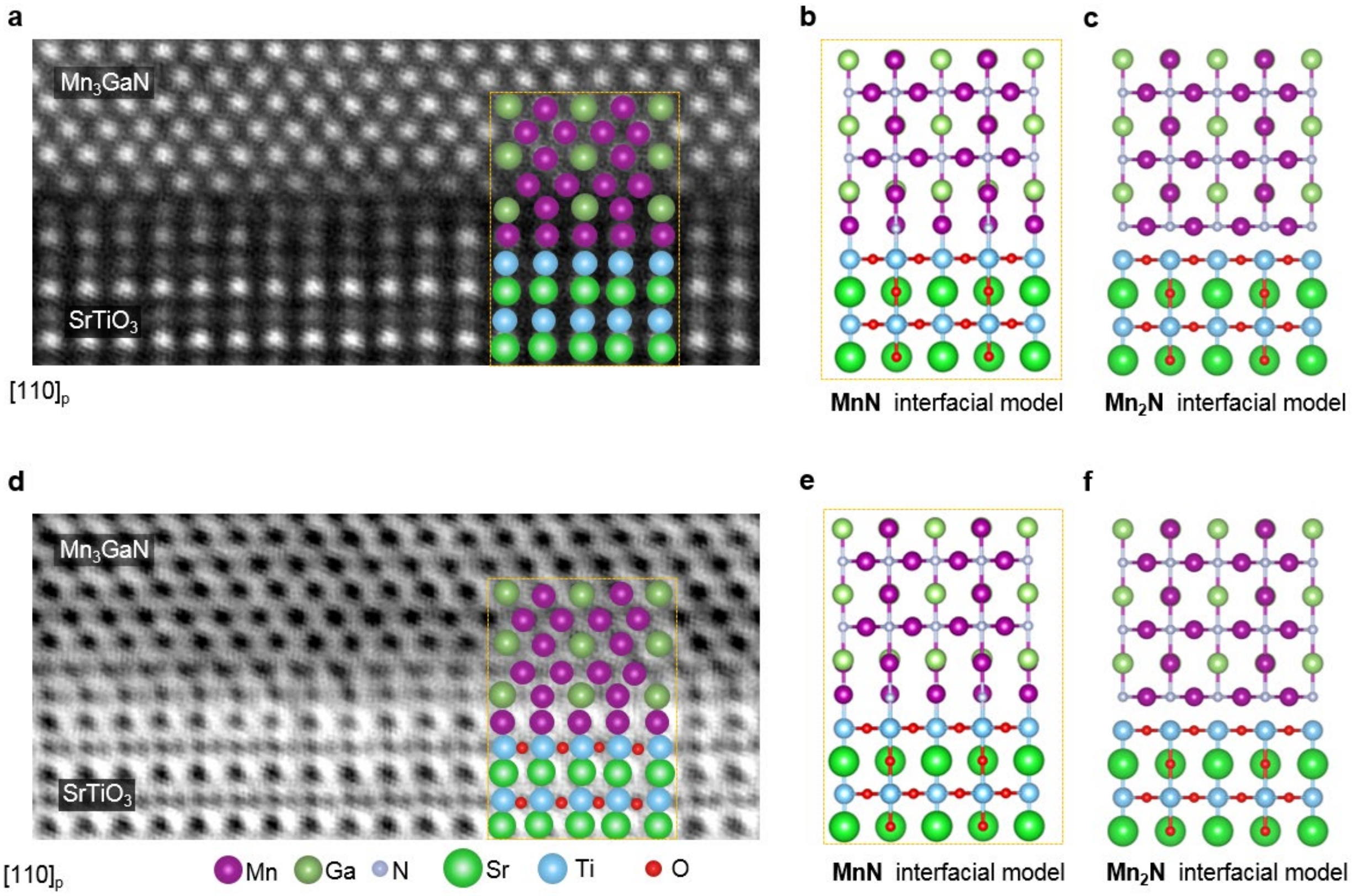

Fig. S9. (**A**) [110]-projected HAADF-STEM image around the $Mn_3GaN/SrTiO_3$ interface. Overlaid on the HAADF-STEM image is the proposed atomic configuration at the interface based on EELS and EDS analyses. (**B**) [110]-projection of the interfacial model considering a MnN interfacial layer. (**C**) [110]-projection of the interfacial model considering a $Mn_2N$ interfacial layer. (**D**) [110]-projected ABF-STEM image around the $Mn_3GaN/SrTiO_3$ interface. Overlaid on the ABF-STEM image is the proposed atomic configuration at the interface, which matches with the interfacial model considering a MnN interfacial layer as shown in e but disagrees with the $Mn_2N$ interfacial model as shown in (**F**).

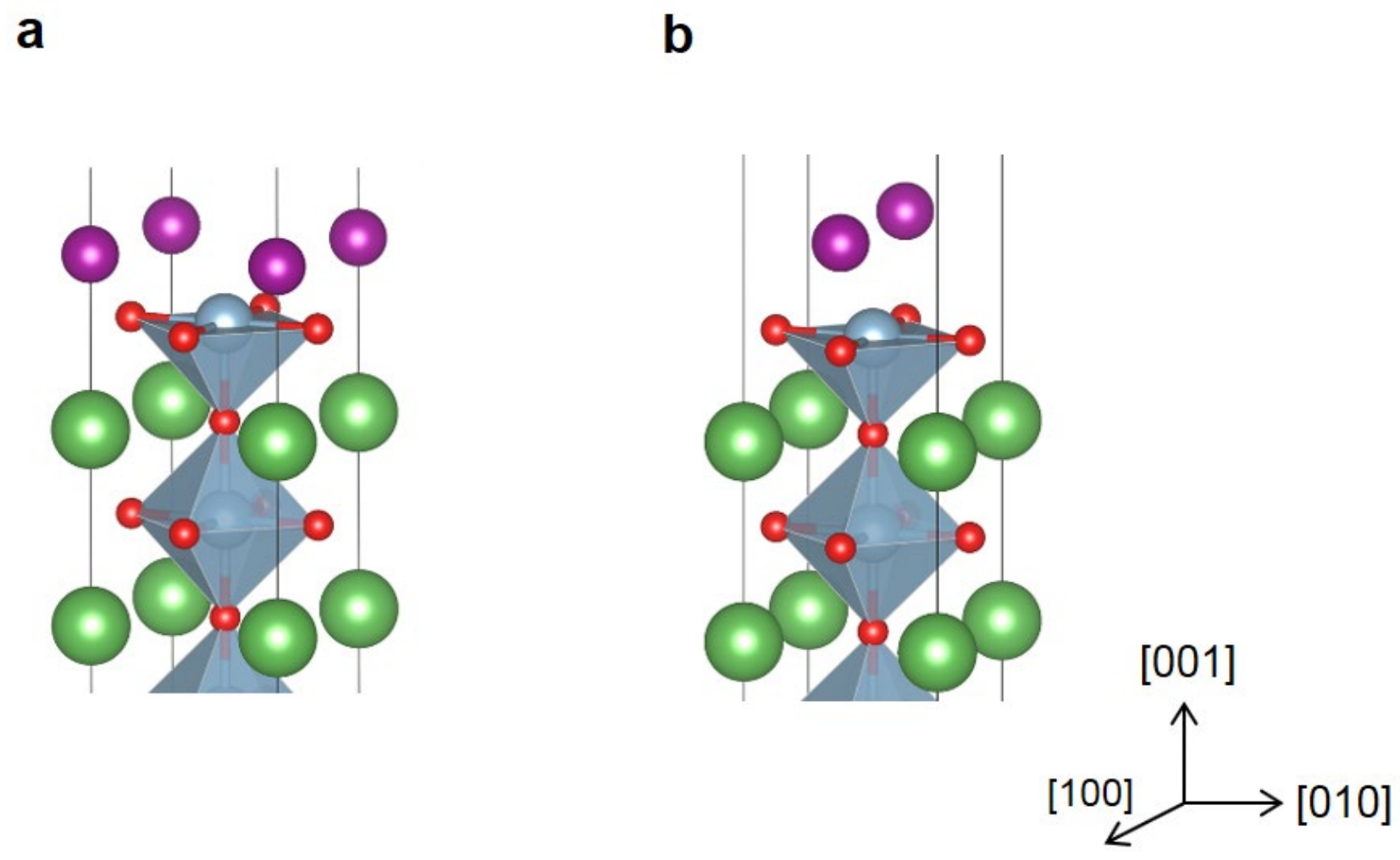

Fig. S10. Representation of the Mn/$ABO_3$ supercells used for the theoretical calculations for two different Mn configurations, (**A**), Mn(1) and (**B**), Mn(2), as described in Figure 5.

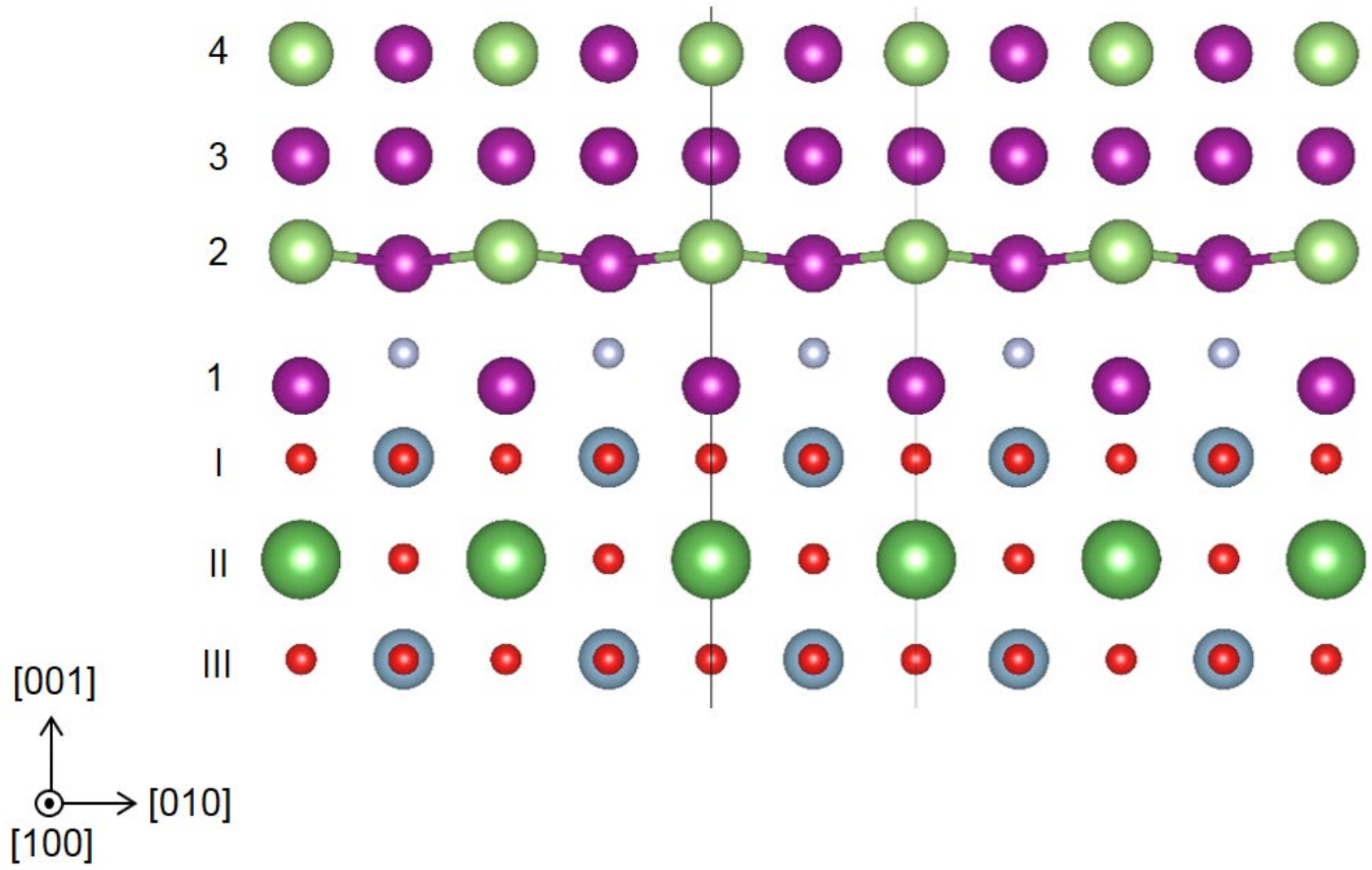

Fig. S11. The puckered experimentally observed GaMn layer (layer 2) is successfully reproduced in our computational calculations.